\documentclass{article}[12pt]
\usepackage{graphicx}
\newcommand{\url}{\tt}
\makeindex
\begin{document}

\title{\bf Producing an Intense, Cool Muon Beam via {\boldmath $e^+e^-$} Annihilation\thanks{Prepared for Proceedings of the NuFact06 Workshop, Irvine, CA, 23--30 August 2006.}}

\author{
D. M. Kaplan,$^*$ T. Hart\\
{\sl Physics Division, Illinois Institute of Technology,}\\ 
{\sl Chicago, Illinois 60616, USA}\\[.1in]
{$^*$E-mail: kaplan@iit.edu}\\[.2in]
{P. Allport}\\
{\sl University of Liverpool, Liverpool, UK}}
\date{28 November 2006}
\maketitle

\begin{abstract}
We consider a highly unconventional approach to generating muon and antimuon bunches for a neutrino factory or muon collider: electron-positron annihilation just above muon-antimuon threshold. This approach can produce low-emittance bunches at high energy, easing the muon-cooling and acceleration challenges in such facilities. However, the small ($<$\,1\,$\mu$b) useable production cross section means that extraordinary beam-power and targeting challenges would have to be met. We speculate on what this might entail.
\end{abstract}

\section{Introduction}

In principle, low-emittance bunches of muons for a neutrino factory~\cite{Geer} or muon collider~\cite{Ankenbrandt} could be produced by $e^+e^-$ annihilation near $\mu^+\mu^-$  threshold, obviating the need for muon cooling, and, in the case of a neutrino factory, for muon acceleration as well. If this were feasible, one might hope for a significant cost saving compared to the ``conventional" approach to a neutrino factory or muon collider. The desired rate of muon production is $\sim10^{13}$--$10^{14}$ per second~\cite{Alsharoa}. Might this conceivably be feasible?

\section{Cross section and needed luminosity}

\begin{figure}
\centerline{
\includegraphics[width=.6\linewidth
]{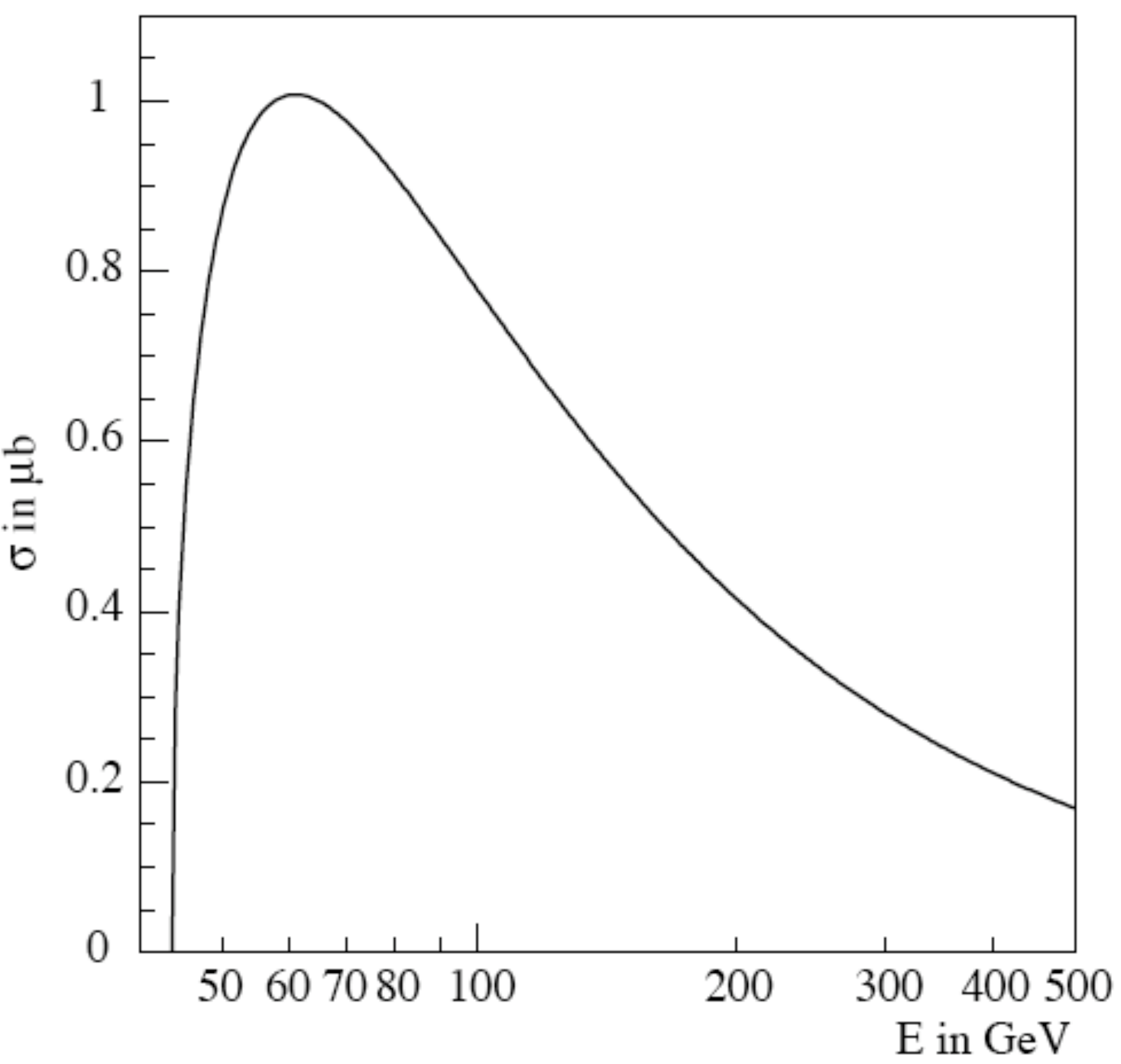}}
\caption{Cross section vs.\ positron energy for $e^+e^-$ annihilation to $\mu^+\mu^-$ assuming a stationary electron target~\protect{\cite{Geant}}.}
\label{fig:sigma}
\end{figure}

The cross section  vs.\ energy for $e^+e^-\to\mu^+\mu^-$ is shown in Fig.~\ref{fig:sigma} for a beam incident on a fixed target~\cite{Geant}. It rises rapidly from a threshold at 43.7\,GeV and peaks at $\approx$\,1\,$\mu$b at an energy of 60\,GeV. To produce a beam of cool muons by this mechanism requires an energy not much above threshold, so that the muons are nearly at rest in the center-of-mass system.
The $<$\,1$\,\mu$b cross section implies a very high luminosity requirement:
\begin{eqnarray*}
{\cal L} &=& n/\sigma \\
&\stackrel{>}{_\sim}&10^{14}\,{\rm s}^{-1}/\,10^{-30}\,{\rm cm}^2\\
&\stackrel{>}{_\sim}&10^{44}\,{\rm cm}^{-2}{\rm s}^{-1}\,.
\end{eqnarray*}
\section{Is there a possible implementation scheme?}

\subsection{Colliding beams?}

The desired center-of-mass collision energy is $\sqrt{s}\stackrel{<}{_\sim} 250$\,MeV; for colliding $\approx$\,125\,MeV electron and positron beams, the above luminosity is clearly far too high to be practical. In any case, since one wants the muons at high energy in the laboratory, colliding-beam production would be undesirable.

\subsection{Fixed-target?}

Positron beams of many amperes are under discussion for a future ``Super {\it B}-Factory"~\cite{SuperB}. If such a beam, of energy $\approx45$\,GeV, could be continuously extracted and made to interact in an electron target of typical liquid or solid density, production of $\stackrel{>}{_\sim}10^{13}$ cool muons/s might then be feasible. 

\label{sec:E158}
However, 
one can quickly rule out conventional fixed-target production such as this on grounds of excessive target-power dissipation. For example, SLAC E-158 operated a 1.5\,m liquid-hydrogen target in a 12\,$\mu$A beam of 48\,GeV electrons. The beam power dissipated in the target was some 500\,W~\cite{E158}, which would scale up to $P_{\rm target}\sim10^8$\,W for interaction rates of interest here. Furthermore, the power in the beam,
\begin{eqnarray*}
P_{\rm beam} \sim 10{\,\rm A}\times10^{11}{\,\rm eV}
= 1{\, \rm TW}\,,
\end{eqnarray*}
would need to be continuously supplied, as well as dissipated in a beam dump. While the high target-power dissipation might be sustainable in a target of radical design~\cite{Blondel-private,Brooks-private}, the sustained provision and safe dumping of such high beam power would be a very substantial challenge.

\subsection{Positron storage ring?}

The preceding argument implies that for any such scheme to be practical, the beam power must be continuously recycled rather than dumped\,---\,for example, in a positron storage ring with internal target. But even a storage ring the size of LEP (for example) would require a prohibitive amount of rf power to compensate the synchrotron-radiation loss. To maintain a 104.5\,GeV beam energy, the LEP rf system provided 3.63\,GeV of acceleration per turn~\cite{LEP}. While this scales down to just $\approx120$\,MeV per turn at 45\,GeV, at a positron current of 10\,A some 13\,TW of rf power would have be continuously supplied to the beam. The $E^4/r$ radius dependence of synchrotron-radiation loss means that making the ring bigger is not a cost-effective solution: a factor $>10^3$ in storage-ring size would be called for, or a circumference $>27\times10^3$\,km.

\subsection{Energy-recovering linac?}
One is thus led to a positron energy-recovering linac (ERL) as perhaps the only way to make this work. The required ERL would be novel for its high energy and high current, as well as for the use of positrons rather than electrons. While none of these is obviously a show-stopper,  a beam power of 250\,kW was the maximum achieved in an ERL as of $\approx\,$2003~\cite{ERL}; MW-scale ERLs are currently under development~\cite{Neiletal,Litvinenkoetal}. Unlike in an electron ERL, the enormous positron current required most likely could not be provided by a positron source in a single pass, but would have to be accumulated. Thus, once accumulated, the positrons themselves would need to be recycled from one pass through the ERL to the next, imposing a stringent limit on emittance growth during the acceleration and deceleration phases.

Given the terawatt beam power, high energy-recovery efficiency is required. At a typical 99.95\% efficiency~\cite{Litvinenkoetal}, of order a gigawatt of rf power would have to be continually supplied to the beam.

\subsection{Target considerations}
\subsubsection{Electron target}
An ideal target for such an application would consist entirely of electrons, reducing the target-power dissipation by some five orders of magnitude relative to the estimate of Sec.~\ref{sec:E158}. A possibly relevant development is the space-charge lens, in which electron densities of order 10$^9 \,e/$cm$^{3}$ have recently been demonstrated~\cite{Pozimski}. However, this is a far cry from the $\stackrel{>}{_\sim}$\,$10^{20} \,e$/cm$^{3}$ that would be needed. Furthermore, in the space-charge lens of Ref.~\cite{Pozimski}, the electron charge-density excess relative to that of negative ions was only about 33\%, thus the problem of large target-power dissipation remains.

It appears unlikely that there is a workable electron-target solution for the desired muon rates. For example, high-current micro-pulse electron guns have been discussed with current densities of 400\,A/cm$^2$ or more~\cite{Makoetal}, and densities some orders of magnitude greater may be feasible~\cite{Mako}. Even with such a source, Table~\ref{tab:example} shows that the production rate misses the goal by  8 orders of magnitude.

\begin{table}
\begin{center}
\caption{Estimates of parameters and performance that might be feasible, assuming a 100\,A positron beam incident on a high-current electron-source target.}\label{tab:example}
\vspace{0.05in}
\begin{tabular}{lccl}
\hline\hline
Parameter & Symbol & Value & Unit \\
\hline\hline
Useable annihilation cross section & $\sigma_{\rm eff}$ & $10^{-31}$ & cm$^2$ \\ \hline
Positron current &&100 & A \\
Positron flux & $N_+$ & $6\times10^{20}$ & s$^{-1}$\\ \hline
Electron-source current density& & 100 & kA/cm$^2$ \\
Electron-target volume density & & $2\times10^{13}$ & $e$/cm$^{3}$\\
Electron-target length along beam && 100 & cm \\
Electron-target areal density& $J_-$ & $2\times10^{15}$ & cm$^{-2}$ \\ \hline\hline
Event rate& $N_+\,J_-\,\sigma_{\rm eff}$ & $1\times10^5$& s$^{-1}$ \\
\hline\hline
\end{tabular}
\end{center}
\end{table}

\subsubsection{Gigawatt target?}

It may be possible to dissipate of order a gigawatt in a water-jet target~\cite{Blondel-private}. Ideally, the resulting steam would be used for co-generation, thereby reducing the load on the power grid of such a facility. The further elucidation of this idea is beyond the scope of this article.

\section{Emittance}

\begin{figure}[tb]
\centerline{\includegraphics*[bb=34 176 560 690,clip,width=.8\linewidth]{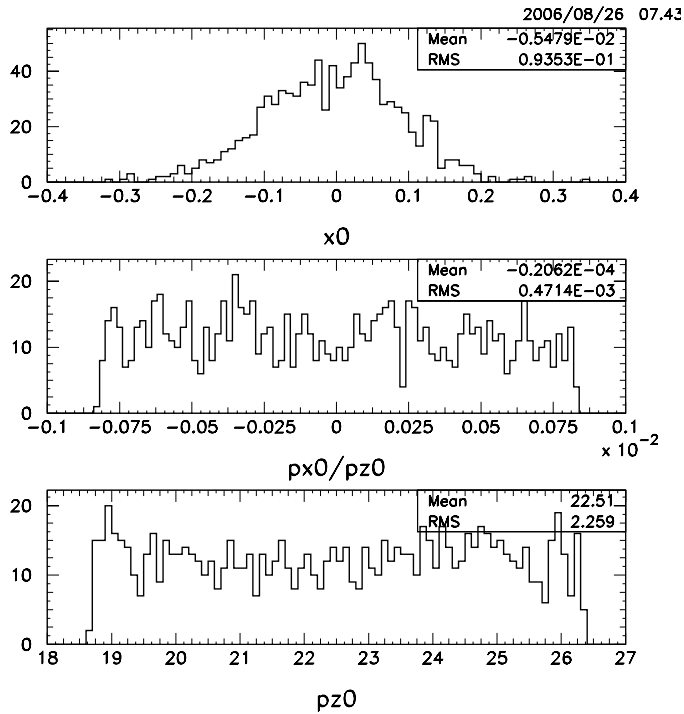}}
\caption{Distributions in horizontal position $x$ (mm), horizontal slope $p_x/p_z$, and longitudinal momentum $p_z$ (GeV/$c$) for a simulated sample of muons produced by a 45\,GeV positron beam incident on a target of stationary electrons; the positron beam was assumed to have a transverse size of 1\,mm and negligible divergence.}\label{fig:emittance}
\end{figure}

The great virtue of an $e^+e^-$-annihilation muon source is that it can produce cool muon beams of both signs. Figure~\ref{fig:emittance} shows muon distributions from a simple Monte Carlo calculation at a positron energy of 45\,GeV. The rms normalized transverse emittance is given by 
\begin{eqnarray*}
\epsilon_n = \gamma\beta\sigma_x\sigma_{x^\prime} = 10\,\pi\,{\rm mm\!\cdot\!mrad}\,,
\end{eqnarray*}
which is compatible with the acceptance of a typical neutrino factory storage ring~\cite{Alsharoa}. However, the momentum spread $\Delta p/p=10$\% is about an order of magnitude too large. A cut $|\Delta p/p|<0.01$ keeps 10\% of the muons, giving an effective production cross section $\sigma_{\rm eff}\approx 0.05\,\mu$b. Since in this reaction muon longitudinal momentum $p_z$ correlates with transverse slope $p_x/p_z$, such a cut also reduces the transverse emittance of the accepted beam by an order of magnitude.

\section{Conclusions}

If electron-positron annihilation to muon pairs is to be a practical source for a muon collider or neutrino factory, a positron energy-recovering linac with  $\approx$\,45\,GeV energy and  $\gg$100\,A current is required. It is worth noting that a muon collider provides luminosity inversely proportional to the beam emittance, whereas the performance figure-of-merit for a neutrino factory depends only on the muon flux. Thus with sufficiently cool muon beams, a collider may be feasible with about an order of magnitude fewer muons per second than are required for a neutrino factory~\cite{LEMC}. 

We note that another scheme to produce muon pairs from an electron beam has also been considered: bremsstrahlung by a high-energy ($\approx$\,50\,GeV) electron beam in a high-$Z$ target. As of 1994, Barletta and Sessler concluded that a muon collider employing such a muon-production scheme would be capable of  $\sim$\,10$^{11}\,\mu$/s and luminosity in the range $10^{27}$--$10^{30}$\,cm$^{-2}$s$^{-1}$, depending on how aggressively one extrapolated from then-current technology~\cite{Barletta-Sessler}.

In conclusion, while it is perhaps premature to rule out an $e^+e^-$-annihi\-lation muon source at this stage, to go further  would certainly require considerable R\&D (some of which is already in progress~\cite{ERL,Neiletal,Makoetal}). Given the extraordinary beam and target parameters required, the cost effectiveness of this approach (compared to the ``conventional" one of pion hadroproduction / decay / muon cooling and acceleration)  is far from clear.

\section{Acknowledgments}
We thank F. Mako, D. Neuffer, J. Pozimski, and D. Summers for useful conversations. This work was supported  by the US Dept.\ of Energy and the National Science Foundation.

\end{document}